\documentclass[useAMS,usenatbib]{mn2e}
\usepackage{epsfig}
\usepackage{psfig}
\usepackage{lscape}

\title[\emph{Fermi} observations of HESS J1745-303 \& HESS J1741-302]{Observing two dark accelerators around the Galactic Centre with \emph{Fermi} Large Area Telescope}

\author[Hui et al.]{C. Y. Hui$^{1}$\thanks{E-mail: cyhui@cnu.ac.kr}, P. K. H. Yeung$^{1,2,3}$, C. W. Ng$^{2}$,
L.~C.~C.~Lin$^{4}$, P. H. T. Tam$^{5}$,  
\newauthor K. S. Cheng$^{2}$, A. K. H. Kong$^{3}$, D. O. Chernyshov$^{6,2}$, and V. A. Dogiel$^{6,2,7}$
\\
$^{1}$Department of Astronomy and Space Science, Chungnam
National University, Daejeon 305-764, Korea\\
$^{2}$Department of Physics, University of Hong Kong, Pokfulam Road, Hong Kong\\
$^{3}$Institute of Astronomy and Department of Physics, National Tsing Hua University, Hsinchu, Taiwan\\
$^{4}$Institute of Astronomy and Astrophysics, Academia Sinica, Taiwan\\
$^{5}$School of Physics \& Astronomy, Sun Yat-Sen University, Guangzhou 510275, China\\
$^{6}$I.E.Tamm Theoretical Physics Division of P.N.Lebedev Institute of Physics, Leninskii pr.
53, 119991 Moscow, Russia\\
$^{7}$Moscow Institute of Physics and Technology, 141700 Moscow Region, Dolgoprudnii, Russia\\}

\begin{document}

\date{Received 2015 November 5}

\pagerange{\pageref{firstpage}--\pageref{lastpage}} \pubyear{2015}

\maketitle

\label{firstpage}

\begin{abstract}
We report the results from a detailed $\gamma-$ray investigation in the field of two ``dark accelerators", 
HESS~J1745-303 and HESS~J1741-302, with 
$6.9$~years of data obtained by the \emph{Fermi} Large Area Telescope. For HESS~J1745-303, we found that its MeV-GeV emission is mainly 
originated from the ``Region A" of the TeV feature. Its $\gamma-$ray spectrum can be modeled with a single power-law with a 
photon index of $\Gamma\sim2.5$ from few hundreds MeV to TeV. 
Moreover, an elongated feature, which extends from ``Region A" toward northwest for $\sim1.3^{\circ}$, 
is discovered for the first time. The orientation of this feature is similar to that of a large scale atomic/molecular gas distribution.
For HESS~J1741-302, our analysis does not yield any MeV-GeV counterpart for this unidentified TeV 
source. On the other hand, we have detected a new point source, \emph{Fermi}~J1740.1-3013, serendipitously. Its spectrum is 
apparently curved which resembles that of a $\gamma-$ray pulsar. This makes it possibly associated with PSR~B1737-20 or
PSR~J1739-3023. 
\end{abstract}


\section{INTRODUCTION}
  Very high energy (VHE; $>100$~GeV) surveys of our Galaxy with Imaging Atmospheric Cherenkov Telescopes 
have uncovered a population of $\gamma$-ray sources in the TeV regime (e.g. Aharonian et al. 2005, 2006). 
For Galactic VHE sources, most of them belong to the classes of pulsar wind nebulae (PWNe) or supernova remnants (SNRs).  
The nature of their VHE emission is primarily identified by their counterparts found at lower energies 
(e.g. radio, X-ray). Their TeV-to-X-ray flux ratio, $f_{\rm 1-10 TeV}/f_{\rm 2-10 keV}$ is typically $<2$ 
(cf. Bamba et al. 2007; Matsumoto et al. 2007). 

 On the other hand, there is a number of VHE sources that no X-ray/radio counterpart has been 
found which makes the identification of their nature very challenging. In any case, the detection of VHE 
$\gamma$-rays clearly indicates certain kind(s) of acceleration mechanism(s) is/are taking place in these sources, 
though the details of the processes are unknown. And therefore, these unidentified objects have been coined with 
the name ``dark accelerators".

Most of them are extended and concentrated on the Galactic plane which strongly suggests they belong to our Galaxy.\footnote{http://tevcat.uchicago.edu} 
Also, they are clustered within $\sim30^{\circ}$ around the Galactic centre. The difficulties of finding their 
counterparts in lower energies can be ascribed to the severe absorption/extinction toward the Galactic centre. 
On the other hand, the absorption/extinction is less of an issue in MeV-GeV regime. And therefore, the Large Area Telescope
(LAT) onboard \emph{Fermi} Gamma-ray Space Telescope provides the ideal instrument for constraining the nature of these 
unknown TeV sources at lower energies.

One of the most interesting dark accelerators is HESS J1745-303 (Aharonian et al. 2008; Bamba et al. 2009: Hui et al. 2011). 
It was firstly discovered by the H.E.S.S. Galactic Plane Survey (Aharonian et al. 2006) and was subsequently investigated 
in details with dedicated follow-up observations (Aharonian et al. 2008). The TeV $\gamma$-ray image shows that it consists 
several spatial components (denoted as Regions A, B and C in Aharonian et al. 2008). 
Owing to the lack of spectral variability and the insignificant dip among 
these regions in the existing data, it was argued that they are originated from a single object (Aharonian et al. 2008).
 
Searches for the possible X-ray counterpart of HESS J1745-303 have been conveyed with \emph{XMM-Newton} and \emph{Suzaku} 
(Aharonian et al. 2008; Bamba et al. 2009). None of these observations have resulted in any evidence for non-thermal X-ray 
continuum emission. This imposes a limiting TeV-to-X-ray flux ratio to be $>4$ (Bamba et al. 2009) which is larger 
than the typical value of PWNe and SNRs. 

Despite the non-detection of any X-ray continuum, a possible excess of neutral iron K$\alpha$ line emission is discovered 
which is originated from the brightest spatial component (i.e. Region A) of HESS J1745-303 (Bamba et al. 2009). 
It is suggested to be an X-ray reflection nebula, where X-rays from previous Galactic centre activities were reflected by a 
molecular cloud in that location (Bamba et al. 2009; see Koyama et al. (1996a) for a detailed discussion on X-ray reflection nebulae 
around the Galactic centre.).

Searching for the molecular/atomic gas toward HESS J1745-303 has been carried out by Hayakawa et al. (2012). 
They have confirmed that a molecular cloud has a positional agreement with Region A of HESS J1745-303. 
And it is apparently extended toward northwest. The total mass of interstellar protons including both 
the molecular and atomic gas to be $\sim2\times10^{6}M_{\odot}$ which provides sufficient targets for 
producing the observed $\gamma$-rays through hadronic processes, such as proton-proton collision.

Using the first 29 months of data obtained by \emph{Fermi} LAT, we have detected and examined the GeV emission 
from HESS J1745-303 (Hui et al. 2011). We found that  
a simple power law is sufficient to describe the GeV spectrum with a photon index of $\sim2.6$.
Interestingly, the power-law spectrum inferred in the GeV regime can be connected to that of Region A 
of HESS J1745-303 in $1-10$ TeV without any spectral break (cf. Fig.~3 in Hui et al. 2011). However, 
the spatial resolution of the results reported in our previous work did not allow one to discriminate 
whether the GeV emission is from which spatial component(s) of HESS J1745-303 (cf. Fig.~2 in Hui et al. 2011). 
If such spectrally connected GeV-TeV 
emission can be confirmed to be originated from Region A, together with the X-ray reflection and molecular cloud 
on-site, this can shed light on the cosmic ray density and/or the past activities 
of the Galactic centre (see discussion in Hui et al. 2011, Dogiel et al. 2015). This provides us with the 
motivation to revisit this target with the LAT data with a much longer time span and improved calibrations.

Besides HESS J1745-303, we also investigate another dark accelerator HESS J1741-302 in its neighborhood. 
The VHE emission of HESS J1741-302 was detected by Tibolla et al. (2008). The TeV centriods of these two sources are 
separated by $\sim1^{\circ}$. Similar to HESS J1745-303, the source apparently comprises two separate spatial
components (HESS J1741-302A and HESS J1741-302B). 
However, its spectral properties in TeV regime was poorly constrained (Tibolla et al. 2008, 2009). 
Matsumoto et al. (2010) have claimed to found an X-ray counterpart of one component HESS J1741-302A (see Fig.~2 in Matsumoto et al. 2010) 
with \emph{Suzaku}. Its X-ray spectrum can be fitted by a power-law with a photon index of $\Gamma\sim1.1$. And the 
TeV-to-X-ray flux ratio is found to be $\sim6$. For the field of another spatial component, HESS J1741-302B, 
no X-ray counterpart has been found by \emph{Suzaku} except for a serendipitously detected cataclysmic variable candidate 
(Matsumoto et al. 2010). 

Recently, Hare et al. (2015) have performed a high resolution X-ray search with \emph{Chandra} observations. 
19 X-ray sources have been detected in the fields of HESS J1741-302A and HESS J1741-302B. However, the authors concluded that none 
of these sources are likely to be associated with the VHE emission. The previous X-ray counterpart of HESS J1741-302A claimed by 
Matsumoto et al. (2010) based on the \emph{Suzaku} data is resolved into a number of faint point sources without any diffuse emission in 
the high resolution \emph{Chandra} image.
On the other hand, Hare et al. (2015) have suggested that a nearby energetic pulsar PSR~B1737-30, which locates at a distance of $\sim400$~pc and
with a characteristic age and spin-down power of $\tau\sim20$~kyr and $\dot{E}\sim8\times10^{34}$~erg/s respectively (Manchester et al. 2005), 
can possibly contribute
relativistic particles and give rise to the observed VHE emission. Searching for the possible GeV counterpart in the third \emph{Fermi} point 
source catalog (3FGL; Acero et al. 2015) does not yield any identification (Hare et al. 2015). 

In this investigation, we explore the MeV-GeV emission in the field of both HESS J1745-303 and HESS J1741-302 by using a $\sim7$~years of
\emph{Fermi} LAT data with the latest instrumental responses and background models. 

\section{OBSERVATION \& DATA REDUCTION}
In this analysis, we use the data obtained by LAT between 2008 August 4 and 2015 June 24.  
The data were reduced and analyzed with the aid of \emph{Fermi} Science Tools v10r0p5 package. 
In view of the complicated environment of the Galactic central region, we adopted the events classified as 
Pass8 ``Clean" class for the analysis so as to better suppress the background. 
The corresponding instrument response functions (IRFs) ``P8R2$\_$CLEAN$\_$V6" 
is used in the investigation.

Considering the spectral distortion induced by energy dispersion can be $>$5\% and rises rapidly below 300 MeV, 
and we are investigating a crowded region around our Galactic centre, 
we focused on the photons with energy $>$650 MeV. We found that including the photons with energies $<$650~MeV makes 
the excesses in the regions of HESS~J1745-303 and HESS J1741-302 difficult to resolve. Also, the signal-to-noise
ratio of $\gamma-$ray features concerned in this investigation drop drastically at energies $>25$~GeV. 
With the aforementioned considerations, we focused on the analysis in the energy range of 0.65-25~GeV. 
We further filtered the data by accepting only the good time intervals where 
the region-of-interest (ROI) was observed at a zenith angle less than 90$^\circ$ 
so as to reduce the contamination from the albedo of Earth.
In this work, we only consider a source is genuine if it is detected at a signal-to-noise level $\geq10\sigma$. 

\section{DATA ANALYSIS}
\subsection{Spatial Analysis}

We started with a binned maximum likelihood analysis for a 28$^\circ$$\times$28$^\circ$ ROI centred at 
RA=$17^{h}41^{m}52.514^{s}$, Dec=$-29^\circ51'05.08"$ (J2000), which is approximately the midpoint between HESS~J1745-303 and HESS J1741-302. 
For subtracting the background contribution, we have included 
the Galactic diffuse background (gll$_-$iem$_-$v06.fits), the isotropic background (iso$\_$P8R2$\_$CLEAN$\_$V6$\_$v06.txt) 
as well as all the point sources reported by the third \emph{Fermi} point source catalogue (3FGL; Acero et al. 2015) within 
31$^\circ$ from the ROI centre in the source model. 
We thawed the spectral parameters of the 3FGL sources within 5$^\circ$ from the ROI centre in the 
analysis. For the other 3FGL sources, their spectral parameters were fixed at the catalogue values. 

Since HESS~J1745-303 is cataloged in 3FGL (3FGL J1745.1-3011), it has already included in the source model through 
the aforementioned procedure. For HESS J1741-302, we added it in the model by assuming a point source at an nominal 
position of RA=$17^{h}41^{m}00^{s}$, Dec=$-30^\circ12'00"$ (J2000)$^{1}$ 
with a simple power-law spectrum. With this source model, we performed a preliminary binned likelihood analysis with the aid of the
tool \emph{gtlike}. Using this initial result, we examined the significance of $\gamma-$ray excess around 
these two dark accelerators by computing the test-statistic (TS) map with the tool \emph{gttsmap}. 

\begin{figure*}
\centerline{\psfig{figure=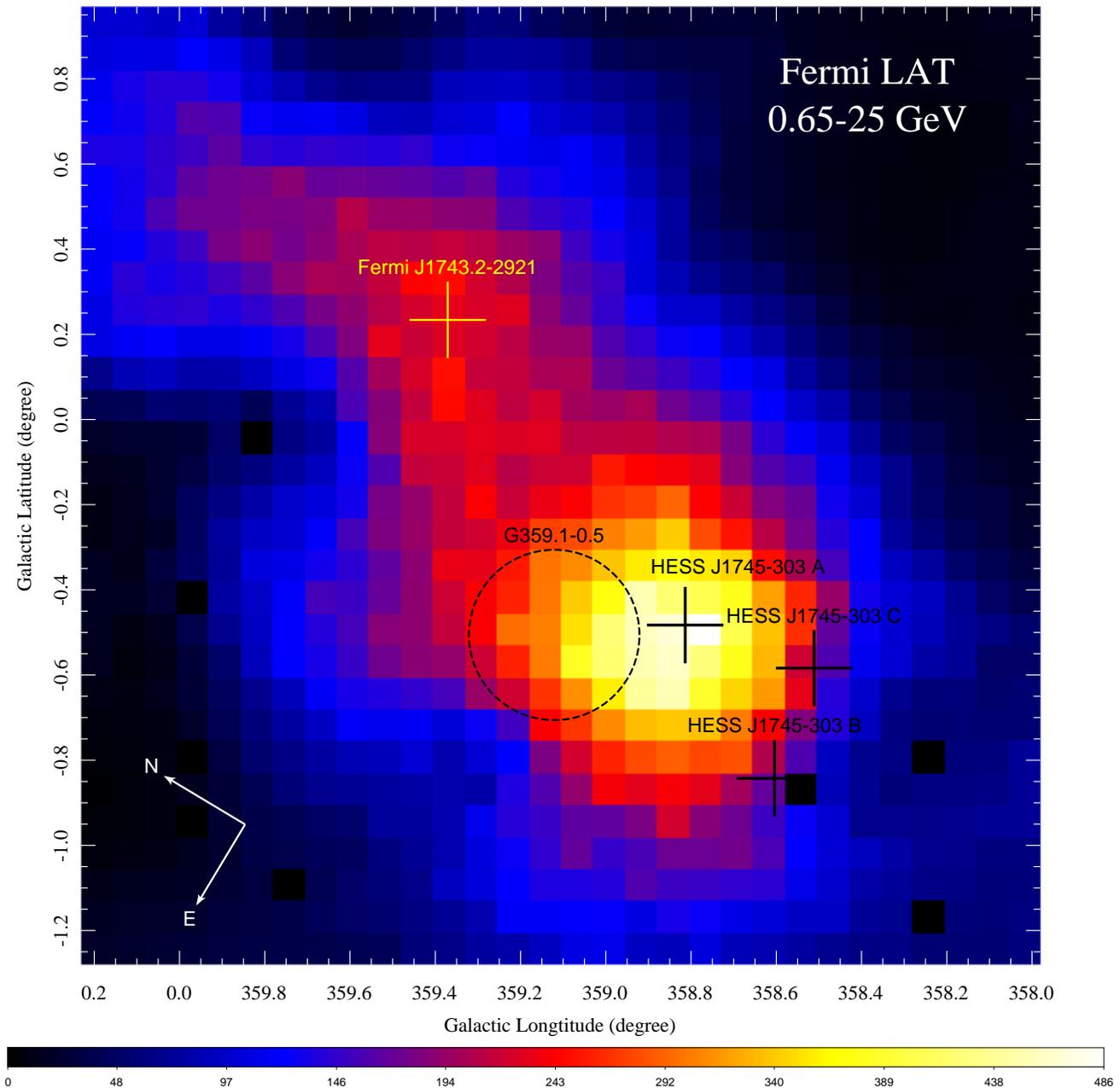,width=18cm,clip=}}
\caption[]{Test-statistic map in 0.65-25 GeV of the field around HESS~J1745-303. 
The nominal emission centres of various TeV components are 
illustrated by the black crosses (see Fig.~1 of Aharonian et al. 2008). The arrows indicate the directions of north and east. 
We note that there is a feature extends toward northwest. 
The location of the clump at the end of the feature, \emph{Fermi}~J1743.2-2921, is 
illustrated by the yellow cross. The location of a nearby supernova remnant G359.1-0.5 with 
a diameter of $\sim24^{'}$ is illustrated by the dashed circle.}
\label{j1745_tsmap}
\end{figure*}

The TS map of the field around HESS~J1745-303 is shown in Figure~\ref{j1745_tsmap}. The $\gamma-$ray emission of this source
can be detected at a significance of $\sim22\sigma$ in 0.65-25 GeV. In comparison with the Figure~2 in Hui et al. (2011), 
we can now resolve the feature much better. It is now clear that most of the MeV-GeV emission of HESS~J1745-303 is 
originated from Region A. Also, an extended feature is detected at $\sim16\sigma$ in this energy range. 
It extends toward northwestern direction for $\sim1.3^{\circ}$. A clump is found at the far end of this feature which  
locates at RA=$17^{h}43^{m}12^{s}$, Dec=$-29^\circ21'00"$ (J2000). For estimating its emission properties, we appended it  
in the source model as a point source with a power-law spectrum in the subsequent analysis. It is referred as 
\emph{Fermi}~J1743.2-2921 hereafter. 
 
We have also investigated the hardness of the $\gamma-$ray emission from HESS~J1745-303 and \emph{Fermi}~J1743.2-2921 
by repeating the aforementioned analysis in soft band $0.65-4$~GeV and hard band $4-25$~GeV. HESS~J1745-303 
is detected at $\sim19\sigma$ (TS=355.7) and $\sim6\sigma$ (TS=40.1) in soft and hard bands respectively. On the other hand, 
\emph{Fermi}~J1743.2-2921 is detected at $\sim13\sigma$ (TS=178.0) in soft band. But its significance drops below $\sim3\sigma$ (TS=7.2) 
in the hard band.

\begin{figure*}
\centerline{\psfig{figure=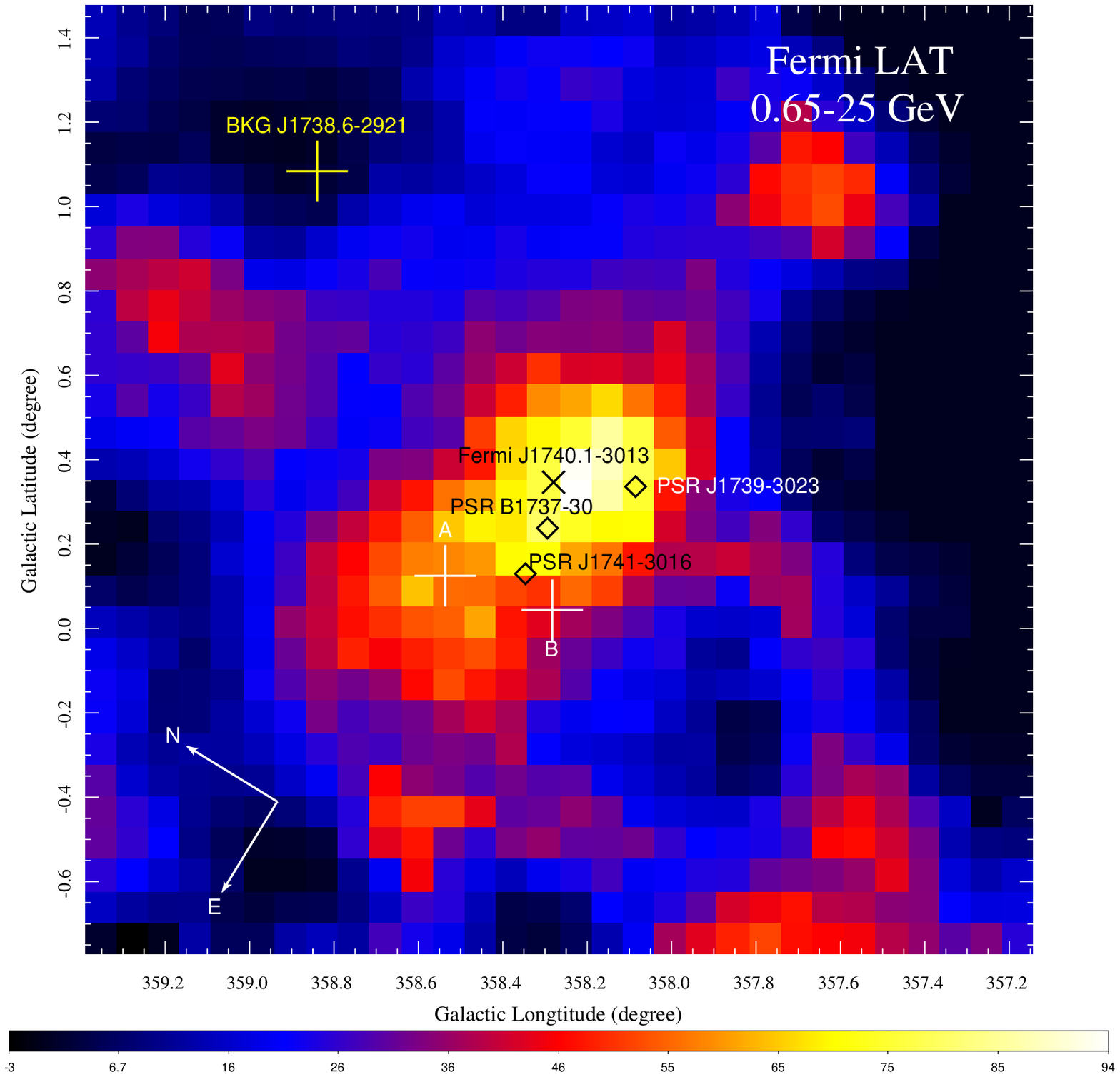,width=18cm,clip=}}
\caption[]{Test-statistic map in 0.65-25 GeV of the field around \emph{Fermi}~J1740.1-3013. 
The nominal emission centres of the TeV components A and B of HESS~J1741-302 (cf. Hare et al. 2015) are illustrated by the 
white crosses. The arrows indicate the directions of north and east. 
Three pulsars that apparently coincide 
with the MeV-GeV excess are illustrated by the black diamonds.
The location of the point-like background component BKG~J1738.6-2921 revealed in this study is illustrated by the yellow cross (see text).} 
\label{j1741_tsmap}
\end{figure*}

Examining the TS map of the field around HESS~J1741-302 from the aforementioned analysis has revealed two clumps of 
$\gamma-$ray excess. We refer the one located at RA=$17^{h}40^{m}06^{s}$, Dec=$-30^\circ13'00"$ (J2000) as \emph{Fermi}~J1740.1-3013. 
Another excess is found at RA=$17^{h}38^{m}36^{s}$, Dec=$-29^\circ21'00"$ (J2000). We noted that the detection significance of this excess is 
$<9\sigma$ over $\sim6.9$~yrs, it is below our pre-defined detection threshold. Therefore, we will not further report its
detailed properties in this work. But it is kept in the source model throughout the analysis as part of the background estimation.
It will be referred as BKG~J1738.6-2921 throughout this work. 

Based on the aforementioned initial results, we have modified our source model accordingly. 
In addition to \emph{Fermi}~J1743.2-2921, \emph{Fermi}~J1740.1-3013 
and BKG~J1738.6-2921 were also appended to the model by assuming a power-law spectrum. On the other hand, the point source component 
of HESS~J1741-302 was removed from the model in view of the non-detection.  

With these modifications, we have re-run the likelihood analysis and produced the TS map 
of the field around \emph{Fermi}~J1740.1-3013 which is displayed as Figure~\ref{j1741_tsmap}. It is detected at 
a signal-to-noise level of $\sim10\sigma$. Its peak is apparently offset from HESS~J1741-302 by $\sim0.4^{\circ}$. 
Although there is a feature extends eastward and towards HESS~J1741-302A, it is only significant at a level of $\sim6\sigma$. 
Therefore, we concluded that \emph{Fermi}~J1740.1-3013 is a newly detected source which is unlikely to be the counterpart of 
HESS~J1741-302. 

We have also analysed \emph{Fermi}~J1740.1-3013 in soft and hard bands. While it remains to be detected at $\sim10\sigma$ (TS=110.2) 
in the soft band which is similar to the signal-to-noise level in the full band analysis, its significance drops below $2\sigma$ (TS=2.2) in 
the hard band.  

\begin{figure*}
\centerline{\psfig{figure=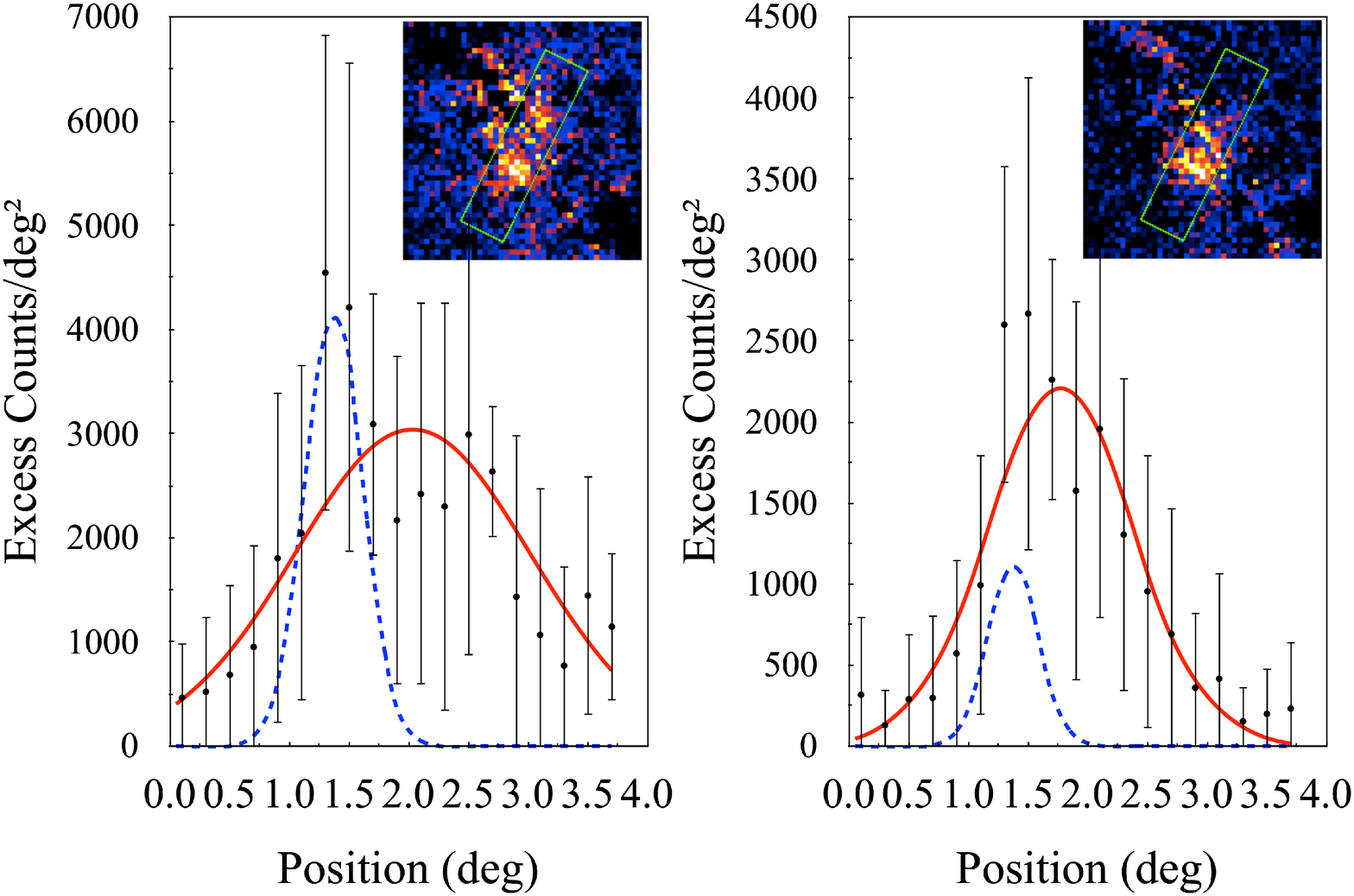,width=18cm,clip=}}
\caption[]{({\it Left panel}): The $\gamma-$ray brightness profile of HESS~J1745-303 computed from the 
background-subtracted count map in 0.65-25~GeV which is displayed as the inset. The box in the inset illustrates 
the orientation that the brightness profile is computed. The red curve shows the best-fit Gaussian model.  
The blue dashed curve illustrates the expected profile of a point-like source. 
({\it Right panel}): Same as left panel but with only the events 
belong to the PSF3 partition selected.} 
\label{j1745_profile}
\end{figure*}

\begin{figure*}
\centerline{\psfig{figure=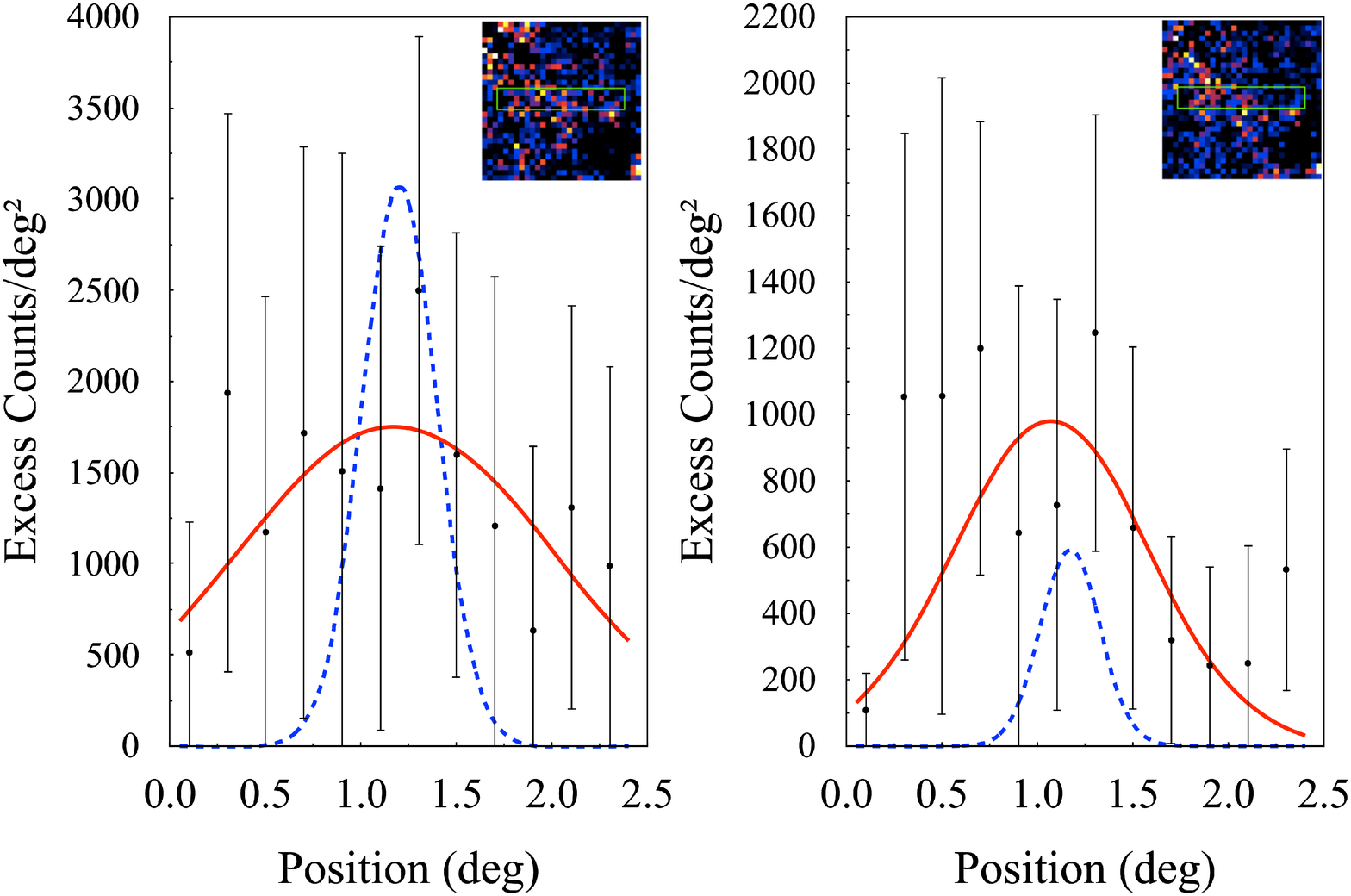,width=18cm,clip=}}
\caption[]{({\it Left panel}): The $\gamma-$ray brightness profile of \emph{Fermi}~J1740.1-3013 computed from the
background-subtracted count map in 0.65-25~GeV which is displayed as the inset. The box in the inset illustrates
the orientation that the brightness profile is computed. The red curve shows the best-fit Gaussian model. 
The blue dashed curve illustrates the expected profile of a point-like source.
({\it Right panel}): Same as left panel but with only the events
belong to the PSF3 partition selected.} 
\label{j1741_profile}
\end{figure*}

For further investigating the MeV-GeV morphology of HESS~J1745-303 and \emph{Fermi}~J1740.1-3013, we produced the 
background-subtracted $\gamma-$ray count-maps of these two sources with a pixel size of $0.1^{\circ}\times0.1^{\circ}$ and computed the brightness profiles 
along the features. The binning factor of the brightness profiles (0.2$^{\circ}$) is chosen with a consideration to optimize the signal-to-noise 
while keeping a reasonable number of data points for model fitting (see below).
The results are shown in the left panels of Figure~\ref{j1745_profile} and 
Figure~\ref{j1741_profile}. 

To examine the source extent, we have fitted each profile with a Gaussian. For 
HESS~J1745-303, it yields a FWHM of $2.3^{\circ}\pm0.3^{\circ}$. 
We have also examined the brightness profile in an orientation orthogonal to the feature. The best-fit 
FWHM is $1.3^{\circ}\pm0.3^{\circ}$. 
All the quoted uncertainties of the brightness profile fitting in this work are 1$\sigma$.

For comparing the observed brightness profile with that of a point source, we have performed simulation 
with the tool {\it gtobssim}. Taking the point-like power-law spectrum 
of HESS~J1745-303 Region A as the input, we have first generated the simulated dataset in 0.65-25 GeV for the same integration time 
and IRFs as the observed data. We then produce the count map and the brightness profile in the same way as aforementioned. We 
found the FWHM of a point-like source profile is $0.6^{\circ}$. The simulated profile is overlaid in Figure~\ref{j1745_profile}
for comparison. The observed brightness profile of HESS~J1745-303 in the examined orientation differs from that of a 
point source by $>5\sigma$. On the other hand, the difference between the observed profile in the orthogonal direction and 
orthogonal direction is less than $3\sigma$. 

We have also examined if this apparently extended feature is a result of the events' uncertainties in the reconstructed incoming direction 
of photons. 
We did this by re-running the aforementioned analysis but with only the events belong to the ``PSF3" partition selected. 
By sacrificing the photon statistics by a factor of $\sim4$, we produced the background-subtracted count map of the events with the most accurate 
reconstructed incoming direction.\footnote{The isotropic model iso$\_$P8R2$\_$CLEAN$\_$V6$\_$PSF3$\_$v06.txt is used for analyzing PSF3 data.} 
The brightness profile of HESS~J1745-303 obtained from the PSF3 data is displayed in the right panel of Figure~\ref{j1745_profile}.  
The Gaussian fit yields a FWHM of $1.5^{\circ}\pm0.2^{\circ}$. 
The profile in the orthogonal orientation has a best-fit FWHM of $1.1^{\circ}\pm0.2^{\circ}$. 
For comparison, the simulated point-like source profile in PSF3 partition has a FWHM of $0.5^{\circ}$. 
The difference between the observed profile in PSF3 partition and that expected for a point source is $5\sigma$ and $3\sigma$ along the feature 
and its orthogonal orientation respectively. 

For \emph{Fermi}~J1740.1-3013, the Gaussian fit to the brightness profiles of full data (left panel of 
Figure~\ref{j1741_profile}) and the PSF3 data (right panel of Figure~\ref{j1741_profile}) along the east-west orientation yield a FWHM of 
$1.9^{+1.0}_{-0.6}$$^{\circ}$ 
and $1.2^{\circ}\pm0.4^{\circ}$ 
respectively. 
On the other hand, the FWHM of the observed profile in the orthogonal direction is $1.7^{\circ}\pm0.4^{\circ}$ (full data) and 
$1.6^{\circ}\pm0.4^{\circ}$ (PSF3) respectively. The profile along the east-west orientation is consistent with that in the 
north-south orientation within the tolerance of the statistical uncertainties. 

For comparison, we have simulated the expected profile of a point source by assuming a power-law spectrum of \emph{Fermi}~J1740.1-3013. 
The simulated profile has a FWHM of $0.5^{\circ}$ (full data) and $0.4^{\circ}$ (PSF3). 
This only differs from the observed profile by less than $3\sigma$. 
Therefore, there is no conclusive 
evidence for the source to be extended. 

\subsection{Spectral Analysis}
In the aforementioned binned likelihood analysis, we have assumed a simple power-law (PL) spectrum for HESS~J1745-303, 
\emph{Fermi}~J1743.2-2921 and \emph{Fermi}~J1740.1-3013, which yields the photon indices of $\Gamma=2.68\pm0.07$, 
$\Gamma=2.82\pm0.11$ and $\Gamma=2.60\pm0.13$ respectively. Besides the PL model, we have also examined whether an exponential 
cutoff power law model (PLE) or a broken power law (BKPL) can better describe their spectra. 
The detailed of their spectral properties are summarized in Table~1. 

\begin{figure}
\centerline{\psfig{figure=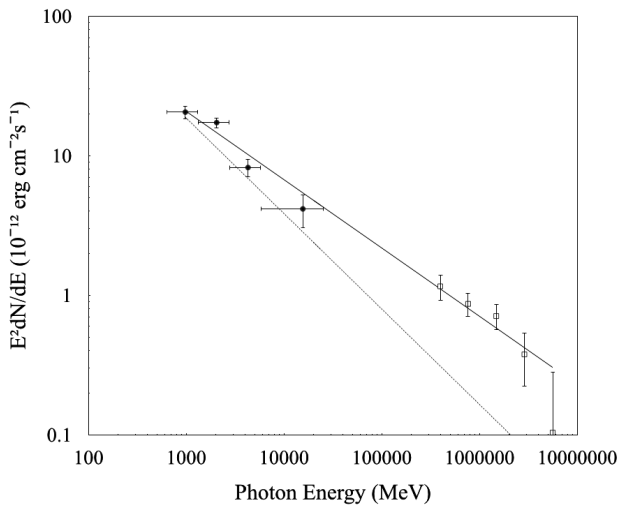,width=9.5cm,clip=}}
\caption[]{Spectral energy distribution of the region A of HESS~J1745-303 as observed by \emph{Fermi} LAT (solid circles) and 
H.E.S.S. (open squares). The dashed line illustrates the PL model ($\Gamma=2.68$) resulted from the likelihood analysis of the 
\emph{Fermi} LAT data. The solid line represents the best-fit PL model ($\Gamma=2.49$) resulted from the 
simultaneously analysis of both \emph{Fermi} and H.E.S.S. data.}
\label{j1745_sed}
\end{figure}

For HESS~J1745-303, the additional spectral parameters in BKPL/PLE are not strongly required based on a likelihood ratio test ($<2\sigma$). 
Therefore, a simple PL model is sufficient to describe its MeV-GeV spectrum. 
For constructing its spectral energy distribution (SED), 
we required each bin to attain a signal-to-noise ratio $>4\sigma$ for a robust result. We started with dividing the 
full energy range into five logarithmically equal spaced energy bins. The binned spectrum is constructed from the independent fitting 
of each spectral bin. Since the significance of the last bin is below our predefined requirement, the last two bins are combined in the SED. 
The resultant SED is shown in Figure~\ref{j1745_sed}.

Hui et al. (2011) have suggested that the GeV spectrum of HESS~J1745-303 can be connected to the TeV spectrum of region A (see Fig.~3 in 
Hui et al. 2011). However, ascribing to the difficulty of resolving the GeV emission in this previous study, this spectral connection 
cannot be confirmed unambiguously. As our spatial analysis shows that the observed MeV-GeV emission is mostly originated from region A 
(see Fig.~\ref{j1745_tsmap}), we can re-examine the spectral connection of the spatial component. 
Fitting the binned spectra obtained by \emph{Fermi} and H.E.S.S. with a single PL model simultaneously yields a 
photon index of $2.49\pm0.02$ with a goodness-of-fit of $\chi^{2}=9.38$ (7 d.o.f.) which is statistically acceptable.

\begin{figure}
\centerline{\psfig{figure=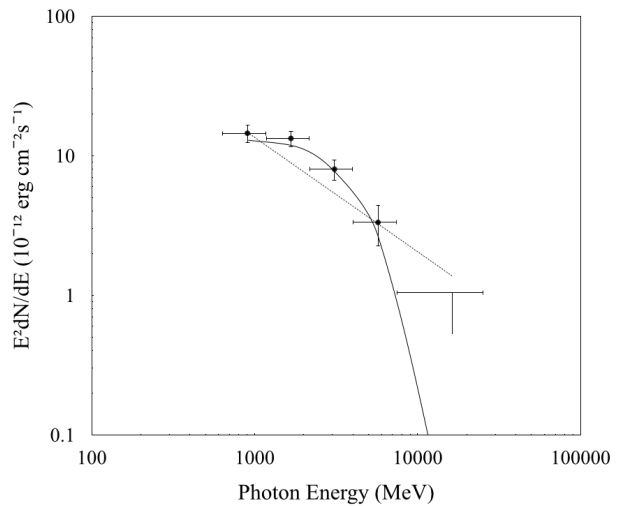,width=9.5cm,clip=}}
\caption[]{Spectral energy distribution of \emph{Fermi}~J1743.2-2921. The solid line and the dotted line illustrate the 
best-fit PLE and PL models respectively. The last bin shows the $2\sigma$ upper-limit. }
\label{j1745_clump_sed}
\end{figure}

For the spectrum of \emph{Fermi}~J1743.2-2921, namely the clump at the end of the 
extended feature associated with HESS~J1745-303, 
the simple PL fit results in a photon index which is consistent with that of HESS~J1745-303 
within the tolerance of $1\sigma$ uncertainties. Its SED is shown in Figure~\ref{j1745_clump_sed}. 
On the other hand, the likelihood ratio test indicates that PLE and BKPL models
are preferred over PL by $>$4$\sigma$. The PLE model yield a photon index of $\Gamma=1.48\pm0.40$ and a cut-off energy 
of $E_{c}=1.90\pm0.67$~GeV. For the BKPL model, we found that the spectral break $E_{b}$ cannot be properly constrained if all 
parameters are taken as free parameters. Therefore, we have fixed it at $E_{b}=2$~GeV for the subsequent analysis. This 
yields the photon indices $\Gamma_{1}=1.94\pm0.24$ and $\Gamma_{2}=3.66\pm0.29$ below and beyond $E_{b}$ respectively. 

\begin{figure}
\centerline{\psfig{figure=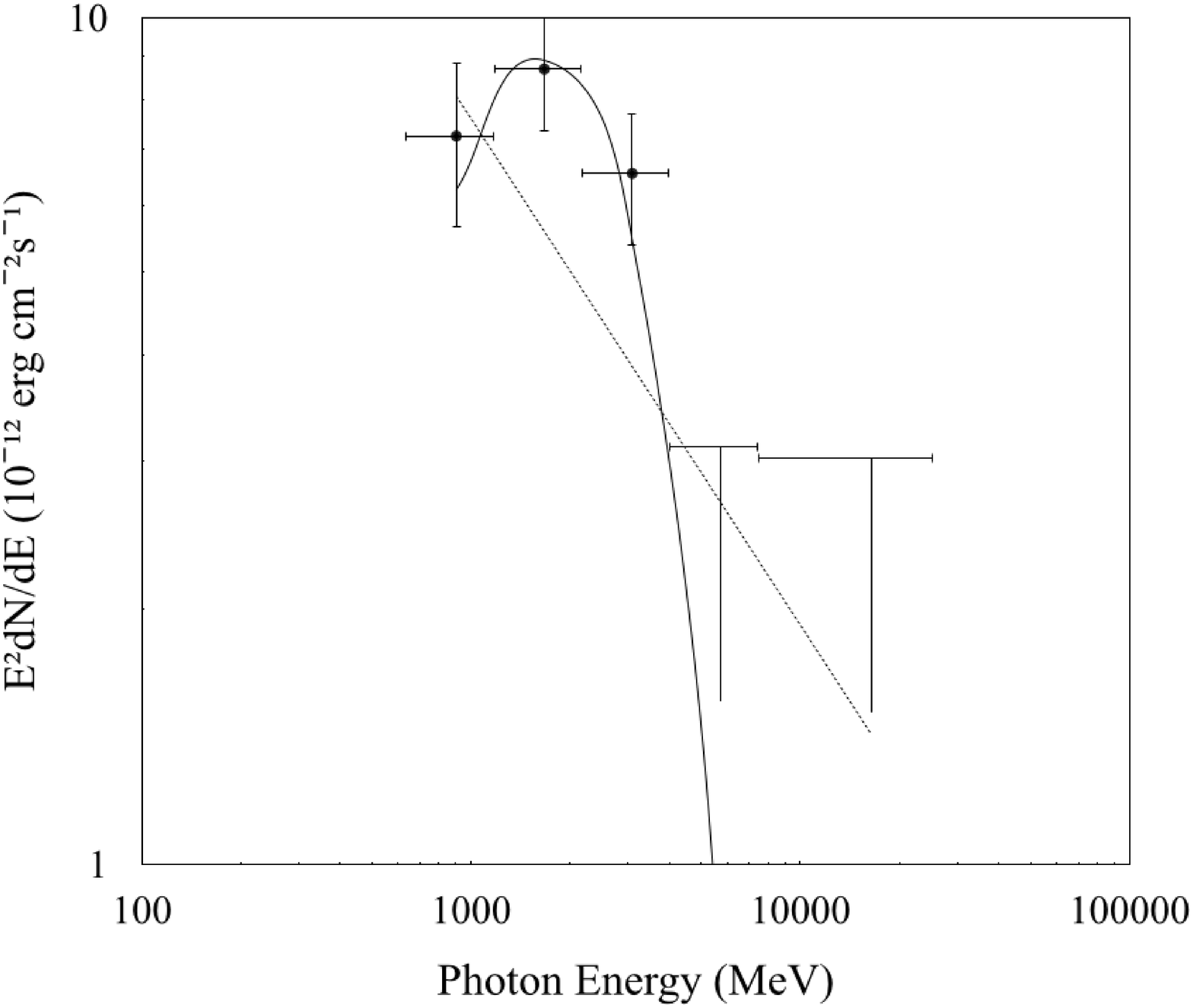,width=9.5cm,clip=}}
\caption[]{Spectral energy distribution of \emph{Fermi}~J1740.1-3013. The solid line and the dotted line illustrate the 
best-fit PLE and PL models respectively. The last two bins show the $2\sigma$ upper limit. }
\label{j1741_sed}
\end{figure}

For \emph{Fermi}~J1740.1-3013, both PLE and BKPL are preferred over simple PL by $>$4.5$\sigma$. 
With the break energy fixed at $E_{b}=2$~GeV, the BKPL fit yields the photon indices of 
$\Gamma_{1}=1.51\pm0.33$ and $\Gamma_{2}=3.67\pm0.42$. For the PLE fit, a photon index of $\Gamma=-0.18\pm0.50$ 
and a cutoff energy of $E_{c}=776\pm164$~MeV is obtained.  
We notice that the inferred $E_{c}$ is very close to the lower bound of our adopted energy range 
(i.e. 650~MeV). To test the robustness of the PLE fit and check if $E_{c}$ is stuck in local optima of the parameter space, 
we repeated the analysis with the lower energy bound shifted to 350 MeV. We found that all the parameters are consistent with 
the previous analysis within the tolerance of statistical uncertainties. The SED of \emph{Fermi}~J1740.1-3013 is shown in 
Figure~\ref{j1741_sed}. 

\begin{table*}
\begin{center}
\caption[]{$\gamma-$ray spectral properties of HESS J1745-303, {\it Fermi} J1743.2-2921 and {\it Fermi} J1740.1-3013 
as observed by \emph{Fermi} LAT.}
\begin{tabular}{lccc}
\\      \hline \hline
      & HESS J1745-303 & {\it Fermi} J1743.2-2921 & {\it Fermi} J1740.1-3013 \\ \hline
\multicolumn{4}{c}{PL}\\ \hline
        $\Gamma$ & 2.68$\pm$0.07 & 2.82$\pm$0.11 & 2.60$\pm$0.13 \\
        Photon flux (10$^{-9}$ photons cm$^{-2}$ s$^{-1}$)$^{a}$ & 14.20$\pm$0.86 & 10.15$\pm$0.90 & 5.89$\pm$0.72 \\ 
        $TS$ & 398.7 & 166.3 & 93.36 \\ \hline 
\multicolumn{4}{c}{PLE}\\ \hline
        $\Gamma$ & 1.95$\pm$0.28 & 1.48$\pm$0.40 & -0.18$\pm$0.50 \\
        $E_c$ (MeV) & 3619.62$\pm$1545.01 & 1897.6$\pm$672.576 & 775.919$\pm$164.014 \\
        Photon flux (10$^{-9}$ photons cm$^{-2}$ s$^{-1}$)$^{a}$ & 13.59$\pm$0.85 & 9.78$\pm$ 0.88 & 5.50$\pm$0.63 \\ 
        $TS$  & 402.67 & 186.78 & 117.26 \\ \hline
\multicolumn{4}{c}{BKPL}\\ \hline
        $\Gamma_1$ & 2.09$\pm$0.15 & 1.94$\pm$0.24 & 1.51$\pm$0.33 \\
        $\Gamma_2$ & 3.18$\pm$0.18 & 3.66$\pm$0.29 & 3.67$\pm$0.42 \\
        $E_b$ (MeV) & 2000 & 2000 & 2000 \\
        Photon flux (10$^{-9}$ photons cm$^{-2}$ s$^{-1}$)$^{a}$ & 13.39$\pm$0.87 & 9.69$\pm$0.88 & 5.63$\pm$0.69 \\ 
        $TS$ & 402.26 & 184.05 & 113.66 \\ \hline
\end{tabular}
\end{center}
\footnotesize{$a$: The energy range of the quoted photon flux in 0.65-25 GeV.} 
\end{table*}

\subsection{Searches for long-term variability}
In order to examine the long-term variability, we divided the $\sim$6.9 years of \emph{Fermi} LAT observation 
into several segments with a constraint that the source of interest can be detected at a level $>3\sigma$ in each 
segment. Binned maximum-likelihood analysis was performed for individual segment by assuming a simple PL 
spectral model for each source of interest. Using the $\gamma-$ray flux obtained from each time bin, 
we construct the light-curves over a time span of $\sim7$~years. 

For HESS~J1745-303 and \emph{Fermi}~J1743.2-2921, the full dataset is split into 11 bins each with a size of 230 days. 
We found that the temporal behavior 
of HESS~J1745-303 and \emph{Fermi}~J1743.2-2921 differs from a uniform distribution at a confidence level 
of $\sim95\%$ and $\sim23\%$ respectively. Hence, there is no evidence of long-term variability for 
\emph{Fermi}~J1743.2-2921. For HESS~J1745-303, the variability is marginal. To further investigate 
its temporal properties, we have repeated the analysis with PSF3 data so as to examine whether the PSF wings of the 
nearby sources might result in the apparent variability of HESS~J1745-303. Adopting the same binning as in the 
full dataset, the confidence level for being a variable drops below $90\%$ in the analysis of PSF3 data. 
Therefore, there is also no solid
evidence for the $\gamma-$rays from HESS~J1745-303 Region A to be variable. 

For \emph{Fermi}~J1740.1-3013, since its detection significance is relatively low, we found it is difficult to split the full 
dataset into a reasonable number of time segments and have each to achieve a signal-to-noise level $>3\sigma$ at the same time. 
Therefore, we are not able to perform any conclusive investigation for its long-term variability in the current study.

\section{ DISCUSSION \& SUMMARY }
We have investigated the $\gamma-$ray emission from the fields of HESS~J1745-303 and HESS~J1741-302 by using $\sim7$ years of data 
obtained by \emph{Fermi} LAT. With the data of improved calibration, we found that the MeV-GeV emission of HESS~J1745-303 
is mainly originated from the Region A (cf. Fig.~\ref{j1745_tsmap}). 
The improved resolution allows us to re-examine and confirm the GeV-TeV spectral connection 
(cf. Fig.~\ref{j1745_sed}) as suggested by our previous study (Hui et al. 2011). 
We note that all the multi-wavelength counterparts, including radio, X-ray and 
GeV, associated with HESS~J1745-303 are from the Region A. This leads us to speculate if the Regions B and C as found by 
Aharonian et al. (2008) are separate sources which are not physically related to HESS~J1745-303. 

In probing the origin of the $\gamma-$ray emission from HESS~J1745-303, a nearby supernova remnant (SNR) G359.1-0.5 has been 
suggested as the contributor of the observed $\gamma-$rays (Aharonian et al. 2008). Through the interactions between the shock
from the SNR and the molecular cloud, the $\gamma-$ray can be produced. 
By modeling the X-ray emission from 
G359.1-0.5, Ohnishi et al. (2011) found that the plasma of this mixed-morphology (MM) SNR is over-ionized. 
This property has also been observed in other $\gamma-$ray emitting MM SNRs such as IC 443 (Yamaguchi et al. 2009) and W49B (Miceli et al. 2010). 
The $\gamma-$ray luminosities of MM SNRs in 0.1-100 GeV lie in the range of $\sim10^{35}-10^{36}$~erg~s$^{-1}$ 
(see Bamba et al. 2015). At a distance of 8.5~kpc, the $\gamma-$ray luminosity of HESS~J1745-303 in this band is $1.1\times10^{36}$~~erg~s$^{-1}$ 
which is marginally consistent with the range of other MM SNRs. 

The $\gamma-$ray spectra of many GeV-detected SNRs are soft and show a break at energies of a few GeV (Acero et al. 2015),
which suggest the escape of high energy particles from the acceleration sites. However, there is no evidence for any spectral break to 
be found in the case of HESS~J1745-303. The difference in the GeV spectral shape has put the association between HESS~J1745-303 and 
G359.1-0.5 in question. 

Since the Region A of HESS~J1745-303 is associated with a molecular cloud, the $\gamma-$ray emission at this location can be 
generated by cosmic rays (CRs) penetrating into the clouds from outside. 
CRs could be provided by supernova remnants from the Galactic disk. These CRs are distributed more or less uniformly throughout 
the Galaxy with some excess in the Galactic centre (about a factor of two in comparison with the local CR density near the Earth) . 

Assuming the observed $\gamma-$rays are originated from the CRs, we estimate the CR density inside the Region A of HESS J1745-303. 
This can be estimated from the $\gamma-$ray emissivity per hydrogen atom in the cloud. Hayakawa et al. (2012) has estimated the mass of the cloud in this 
region as $\sim2\times10^{6}M_{\odot}$. Together with observed $\gamma-$ray flux, this implies a emissivity of 
$\sim4.2\times10^{-6}$~photons~s$^{-1}$~H~atom$^{-1}$. 

It is interesting to compare this emissivity with that of the cloud Sgr~B2 which is at a distance of $\sim100$ pc from the Galactic centre.
Its $\gamma-$ray spectrum in MeV-GeV regime can be described by power-law with $\Gamma\sim2.5$ (Yang et al. 2015), which is comparable with that of 
HESS~J1745-303. However, different from the case of HESS~J1745-303, it cannot be connected to its TeV spectrum without a break 
(see Fig. 3 in Yang et al. 2015). 

The total mass of Sgr~B2 possibly ranges from 6 to 15 million solar masses (see Aharonian et al. 2006). 
According to Yang et al. (2015), the emissivity there is about twice higher than the local value near the Earth. 
This is in a good agreement with predictions of GALPROP model (e.g. Ackermann et al. 2011) which describes the 
average distribution of CRs in the Galaxy emitted by supernova remnants. 
In Table~\ref{j1745_sgrb2}, we compare the emissivities of Sgr~B2 and HESS J1745-303 Region A.
Their comparable emissivities suggest the CRs inside these clouds have common origin 
Their gamma-ray fluxes are produced by CRs with almost the same 
background density in the Galactic centre.

\begin{table*}
\caption{Comparison of the properties of HESS~J1745-303 (Region A) and Sgr~B2.}
\begin{tabular}{lllll}
\hline
Cloud & Flux $> 1$ GeV & Mass & Emissivity & Reference\\
& $10^{-8}$ ph$\cdot$s$^{-1}$cm$^{-2}$ & $10^6$ M$_\odot$ & $10^{-26}$ ph$\cdot$s$^{-1}\cdot$H atom$^{-1}$ & \\
\hline
Sgr B2 & 3.5 & 6 $-$ 15 & 1.4 $-$ 3.5 & Yang et al. (2015)\\
HESS J1745$-$303 & 1.4 & 2 & 4.2 & This work \\
\hline
\end{tabular}
\label{j1745_sgrb2}
\end{table*}

Apart from the bright feature found at the Region A of HESS~J1745-303, we have also discovered an elongated $\gamma-$ray feature that 
extends toward northwestern direction for $\sim1.3^{\circ}$ with a clump \emph{Fermi} J1743.2-2921 locates at the end of this feature
(see Fig.~\ref{j1745_tsmap}). The photon indices of HESS~J1745-303 Region A and \emph{Fermi} J1743.2-2921 inferred from the simple PL 
fits are consistent within the tolerance of statistical uncertainties. A more detailed investigation of their SEDs suggests the 
$\gamma-$ray spectrum of \emph{Fermi} J1743.2-2921 is likely to be more curved. This may indicate the possible spectral steepening 
along the extended feature. 

Since the newly discovered extended feature is apparently connected to Region A, 
it is not unreasonable to speculate that the whole feature has a common origin. To test this hypothesis, we encourage an 
extensive mapping of 6.4 keV line across the whole feature and investigate if the line intensity has any spatial variation. 
Apart from bombardment by the sub-relativistic CRs, the 6.4 keV line can also be generated from a front of hard X-ray photons which 
were emitted by Sgr~A* whose luminosity was much higher in the past than it is at present (Koyama et al. 1996b; Ryu et al. 2013).  
This can result in a time-varying line flux that has been observed in Sgr~B2 (Nobukawa et al. 2014). 
Therefore, it is also interesting to examine if there is any temporal variability of line flux in Region A by comparing with the results of 
Bamba et al. (2009). 

Another crucial information for probing the origin of the extended feature 
is the gas distribution in the $\gamma-$ray emission 
region as this can provide targets for producing the observed $\gamma-$rays through hadronic processes. 
We note that the orientation of the extended $\gamma-$ray feature detected by \emph{Fermi} is 
similar to the intensity distribution of $^{12}$CO and H I as found by Hayakawa et al. (2012) which has one end at the Region A and 
extends to northwest (see Fig.~2 in Hayakawa et al. 2012). This might indicate the association between the molecular/atomic gas distribution 
and the $\gamma-$ray feature. However, the field investigated by Hayakawa et al. (2012) is less than $1^{\circ}\times1^{\circ}$ which 
is not wide enough to cover the whole $\gamma-$ray feature. Follow-up radio spectral imaging that covers the whole field in 
Fig.~\ref{j1745_tsmap} is encouraged for further investigation. 

In searching for the MeV-GeV emission from HESS~J1741-302, we have discovered \emph{Fermi}~J1740.1-3013 which is offset from the TeV 
emission region by $\sim0.4^{\circ}$ (cf. Fig.~\ref{j1741_tsmap}) which makes it unlikely to be associated with HESS~J1741-302. 
On the other hand, three pulsars are found to be coincide with \emph{Fermi}~J1740.1-3013 (see Fig.~\ref{j1741_tsmap}). 
The apparently curved spectrum of \emph{Fermi}~J1740.1-3013 also resembles that of a $\gamma-$ray pulsar, though the spectral 
parameters of the PLE fit cannot be tightly constrained with the current data. 
In the followings, we discuss the possibilities of these pulsars as a counterpart \emph{Fermi}~J1740.1-3013. 

PSR~J1741-3016 has a spin-down power of $\dot{E}\sim5\times10^{31}$~erg~s$^{-1}$ and is located at a distance of $d\sim5$~kpc. 
Assuming it is the counterpart of \emph{Fermi}~J1740.1-3013, 
the $\gamma-$ray luminosity is $L_{\gamma}\sim4.7\times10^{34}f_{\Omega}$~erg~s$^{-1}$ at energies $>100$~MeV, 
where $f_{\Omega}$ is the solid angle of the beam divided by 4$\pi$. Since $L_{\gamma}$ exceeds $\dot{E}$ by several orders of magnitude, 
it is safe to exclude PSR~J1741-3016 as the counterpart of \emph{Fermi}~J1740.1-3013. 

Assuming PSR~J1739-3023 is related to \emph{Fermi}~J1740.1-3013, with $\dot{E}\sim3\times10^{35}$~erg~s$^{-1}$ 
and $d\sim3.4$~kpc of the pulsar, a $\gamma-$ray conversion efficiency $L_{\gamma}/\dot{E}\sim0.07f_{\Omega}$ is implied. 
This is found to be typical for a $\gamma-$ray pulsar (cf. Abdo et al. 2013) which makes PSR~J1739-3023 as a possible 
counterpart of this newly detected $\gamma-$ray source. 

Lastly, we consider the possible association with the nearby pulsar PSR~B1737-30 which has $\dot{E}\sim8\times10^{34}$~erg~s$^{-1}$ 
and $d\sim400$~pc. This implies a conversion efficiency of $L_{\gamma}/\dot{E}\sim0.004f_{\Omega}$. Therefore, from an energetic 
point of view, PSR~B1737-30 is also a possible counterpart of \emph{Fermi}~J1740.1-3013. 

To further investigate the physical origin of \emph{Fermi}~J1740.1-3013, high resolution X-ray imaging of its $\gamma-$ray emission
region is encouraged to look for the X-ray counterparts. This can enable one to compare their X-ray positions with the radio timing 
position of the aforementioned pulsars. Since the X-ray position can possibly be constrained to sub-arcsecond accuracy, this can 
certainly facilitate the pulsation search in $\gamma-$ray which can provide the most crucial information for nailing down the nature 
of this $\gamma-$ray source. 

\section*{Acknowledgments}
CYH is supported by the National Research Foundation of Korea through grant 2014R1A1A2058590.
PHT is supported by the One Hundred Talents Program of the Sun Yat-Sen University.
KSC is supported by a 2014 GRF grant of Hong Kong Government under HKU 17300814P.
AKHK is supported by the Ministry of Science and Technology of Taiwan
through grant 103-2628-M-007-003-MY3.
DOC and VAD acknowledge a partial support from the RFFI grants 15-52-52004, 15-02-02358. 
CYH, PHT, KSC, AKHK, DOC and VAD acknowledge support from the International Space Science
Institute-Beijing to the International Team ``New Approach to
Active Processes in Central Regions of Galaxies."
The author would also like to thank Dr. Rui-zhi Yang for the discussion of Sgr B2. 


\bsp

\label{lastpage}
\end{document}